\documentclass[prd, aps, nobibnotes, nofootinbib,
eqsecnum,
preprint,
amsmath,
amssymb]{revtex4-2}

\usepackage{graphicx}
\usepackage{dcolumn}
\usepackage{bm}
\usepackage{hyperref}
\usepackage{mathrsfs}
\usepackage{xcolor}
\usepackage{slashed}
\usepackage{subfigure}

\def\D{{\mathscr D}}

\begin{document}

\title{Spin gauge theory, duality and fermion pairing}
\author{Shantonu Mukherjee}
\email{shantonumukherjee@bose.res.in}
\affiliation{S N Bose National Centre for Basic Sciences, Block JD, Sector III, Salt Lake, Kolkata 700106, India}
\author{Amitabha Lahiri}
\email{amitabha@bose.res.in}
\affiliation{S N Bose National Centre for Basic Sciences, Block JD, Sector III, Salt Lake, Kolkata 700106, India}

\begin{abstract}
We apply duality transformation to the Abelian Higgs model in 3+1 dimensions in the presence of electrons 
coupled to the gauge field. The Higgs field is in the symmetry broken phase, where flux strings can form.
Dualization brings in an antisymmetric tensor potential $B_{\mu\nu}$\,, which couples to the electrons 
through a nonlocal interaction which can be interpreted as a coupling to the spin current. It also couples
to the string worldsheet and gives rise to a string Higgs mechanism via the condensation of flux strings. 
In the phase where the $B_{\mu\nu}$ field is massless, the nonlocal interaction implies a linearly rising 
attractive force between the electrons, which can be interpreted as the result of a pair of strings joining 
the electrons.

\end{abstract}
\date{\today}

\maketitle

\section{Introduction}
Dualities provide a powerful tool to understand phenomena which are not tractable by perturbation techniques. 
Dualities usually relate strongly coupled sector in one theory to the weakly coupled sector in another theory,
in particle physics, string theory, statistical physics and also condensed matter physics. One class of duality 
transformations in quantum field theory involves exchanging a differential $p$-form $A_p$ in
$D$-dimensions with a $D-p-2$-form $A_{D-p-2}$ by the Hodge duality of their exterior derivatives, 
${\rm d}A_p ={} ^*{\rm d}A_{D-p-2}\,.$ For free fields in topologically trivial spacetimes, both are equivalent 
descriptions, but interactions or topological obstructions can break this duality. Dualities in interacting 
theories, when they can be constructed, can lead to deep mathematical and physical insights. In two and three 
dimensions, (anti)self-dual configurations of gauge fields interacting with scalars correspond to solitons. 
In four dimensions, where the dual of a 1-form is also a 1-form, the (anti)self-dual configurations of 
Yang-Mills gauge fields minimize the action of instantons. A particularly interesting duality in four dimensions
is that between a scalar and a two-form, which persists when there is a gauge field coupled to the scalar. 
The scalar field is now compact and the dual two-form appears in a topological $B\wedge F$ interaction 
with the gauge field. If there is no other field in the theory, this is a duality between the strongly 
coupled Abelian Higgs model and topological mass generation mechanism in four dimensions.

The low energy, long wavelength properties of a condensed matter system can be captured in an effective field
theory, Ginzburg-Landau theory being the original example. Effective field theories which describe topological 
states of quantum matter are topological field theories, or more generally, quantum field theories which include 
topological interaction terms. The application of topological quantum field theories to condensed matter systems 
has, in recent years, greatly improved our understanding of both~\cite{Qi:RMP83, Chiu:RMP88, Fradkin:CUP2013, 
	Chan:2012nb, Zhang:1988wy, Zhang:1992eu, Diamantini:2005dj, Diamantini:2011ck}. 
Although the two-form gauge field is ubiquitous in these theories 
of topological matter, the difficulties of coupling it to electrons has stood in the way of a deeper 
understanding of the field and its applicability in condensed matter physics. 

%
%
%
%

This is most easily seen from the point of view of gauge symmetries. 
Vector gauge transformations, which can be called the fundamental or defining
symmetry of the gauge theory of two-form fields, generalize the gauge transformations of electromagnetism to 
$B \to B + {\rm d}\beta$\,, under which the field strength 
$H = {\rm d} B\,$ remains unchanged. 
But unlike the U(1) symmetry of ordinary gauge theory, this appears to 
have no representation as a local unitary transformation of fermions, and the interaction above does 
not remain invariant under this transformation. 

There is also a problem of nonlocality.
Duality between $B$ and vector fields, for the example of the Abelian Higgs model, appears only through the field strength 
as $H = {}^*(A + {\rm d}\phi)\,.$
So one might be encouraged to try an interaction with fermions in the form
$H\wedge {}^*j\,$ where $j$ could be either the usual fermionic current or the axial current.
But this is also not correct. The reason is that the duality relation, between $B$ on one hand and $(A, \phi)$ on the other, 
is not local. Thus in presence of fermions, the actions obtained by replacing one set of fields with the other are not equivalent and
do not lead to the same equations of motion. The nonlocality can be understood if we remember 
that extended objects
rather than point particles are essential to the definition of higher form gauge fields. For example, the two-form $B$ couples 
to worldsheets of strings rather than worldlines of particles, so one expects that the interaction between
$B$ and fermionic fields representing point particles could involve nonlocality in some form.

Such an interaction was proposed in~\cite{Choudhury:2015rua} between a two-form and a 
nonlocal pseudotensor current $J$ related to the curl of the fermion current
\begin{equation}\label{2-current}
J = m \,{}^*{\rm d}\,  \left({\Box}^{-1}\,J_{\psi} \right)\equiv 
m  {}^*{\rm d}\, \int d^4 y \, G(x, y)\,J_{\psi}\,, 
\end{equation}
where $J_{\psi}$ is the electron current, $m$ is a mass scale appropriate to 
the problem where this interaction might be relevant,
and $G(x, y)$ is the Green function of the wave equation, 
\begin{equation}\label{Gxydef}
\Box G(x, y) = \delta^4 (x - y)\,.
\end{equation}
This current is identically conserved, ${\rm d}{}^*J = 0$\,. The velocity field of the
Dirac fermion is given by $v^i = \psi^\dagger\alpha^i \psi$\,, so the conserved charges 
$\Box J^{0i} = -m\epsilon^{ijk}\partial_j v_k$ can be identified with the vorticity field of the 
fermion. There is another way of looking at the conserved charges. 
In the absence of interactions, and in the non-relativistic limit in which the lower components of the
Dirac fermions may be neglected for energies small compared to their mass, the static charge
of the pseudovector current takes a simple form,
\begin{equation}\label{magmom-density}
\left( J^{0i} \right)_{\rm NR} \propto {1\over 2}  \left( \psi^{\dagger} \sigma^i \psi \right) \,.
\end{equation}
The quantity on the right hand side is the spin density of the electron field, or the 
intrinsic magnetic moment density because it is multiplied by the electron charge. 
The proportionality becomes equality if $m$ is chosen to be the 
electron mass. Just as the interaction between charges and currents is mediated by 
the 1-form gauge field $A$\,, the interaction between magnetic moments and their currents 
is mediated by the 2-form gauge field $B$\,. 

We can write an action for the electrons and the gauge fields incorporating this interaction,
\begin{equation}\label{spin-gauge-action}
S = \int \left[\bar \psi\left(i \slashed\partial + e\slashed A \right) \psi - m \bar \psi \psi
+  {g} B\wedge{}^* J - {1\over 2 } F\wedge{}^*F + {1\over 2} H\wedge{}^*H \right]\,.
\end{equation}
This action is invariant under the vector gauge transformations $B \to B + {\rm d}\beta$\,. 
We should think of this as a low energy 
effective action, valid for energy scales well below some cutoff $\Lambda$\,. The number of
degrees of freedom can be worked out by first making the action local using Lagrange multipliers.
The gauge field $A$ has two degrees of freedom as usual, while $B$ carries only one, since
$B_{0i}$ are non-dynamical and the vector gauge symmetry takes away two more degrees of freedom 
(not three, since $\beta$ and $\beta + {\rm d}\chi$ are equivalent gauge parameters).
There are two additional degrees of freedom in the Lagrange multipliers which stay around and 
may be thought of as auxiliary field degrees required for a local formulation of the non-local
spin gauge interaction~\cite{Choudhury:2015rua}. 
By formally treating the nonlocal interaction like any other coupling term, we 
can integrate out the fermions to find that the one-loop action contains
a  $B\wedge F$ interaction.

Such a term corresponds to topologically massive gauge theory, so the potential between 
two sources interacting via the gauge field ought to be Yukawa in the non-relativistic limit. But that is not what happens in this case. 
Calculated directly from the action in Eq.~(\ref{spin-gauge-action}), the potential has two components. One is an ${r}^{-1}$ Coulomb potential which corresponds to a massless gauge field, and the other is a linear potential which is attractive irrespective of the charges or spin alignment of the fermionic sources~\cite{Chatterjee:2016liu}. In this paper we argue that the action of Eq.~(\ref{spin-gauge-action}) can arise in a system described by local fields. 

Since a linear potential is produced by a string, we start with a system which contains string-like objects as well as fermions. As we will see below, the nonlocal interaction appears naturally if we consider Abrikosov-Nielsen-Olesen (ANO) vortex strings in the Abelian Higgs model interacting with charged fermions. 
Such a system, called a boson-fermion model, was originally proposed as a model of high-$T_c$ superconductivity~\cite{Friedberg:1989ji,Friedberg:1989gj,Friedberg:1990eg, Ranninger:1995PRL, Ranninger:1995, Piegari:2003, Micnas:2002, Pawlowski:2010, SALAS201637, Geshkenbein:1997, RevModPhys.66.1125}, but also appear in models of superconductor-insulator transitions~\cite{Cuoco:2004, Dubi:2007, Loh:2016}, of BCS-BEC crossover~\cite{Deng:2007,  Maska:2017}, and of charged Bose liquids~\cite{Kabanov:2005}. The system we consider has certain differences with the boson-fermion model, in particular we do not consider the $\phi\bar{\psi}\psi$ Yukawa interaction of the fermions with the scalar Higgs field. However for ease of convenience we will refer to the system of Abelian Higgs model with charged fermions as a boson-fermion model. These fermions are ``unpaired'' or ``itinerant'', meaning that they are not part of Cooper pairs whose condensation is responsible for the superconducting transition. 
{The main result of our paper is that these fermions show a kind of confinement being joined by a pair of flux strings to other fermions. The path to that argument requires that we bring together several results, not all of them well known.}

We start by showing in Sec.~\ref{dual-string} how the nonlocal interaction between fermions and the 2-form gauge field $B$ arises in the dual picture of ANO strings interacting with fermions via the electromagnetic gauge field. However, it turns out that the resulting action is not exactly what we are looking for. In Sec.~\ref{effpot} we calculate the static potential between two fermions and find that it also is not what we were hoping for, a mass term for the $B$ field prevents the appearance of a linear potential. We show in Sec.~\ref{condensate} that if the strings condense in a kind of Higgs mechanism, then in a phase where the $B$ field is massless, we recover Eq.~(\ref{spin-gauge-action}) and thus the fermions are bound by a linear potential. We end with a discussion on the physical implications of our results.

\section{ANO string, dual variables, and currents}~\label{dual-string}
In the dual formulation of ANO strings in the Abelian Higgs model, the phase of the scalar field is written as the sum of singular and regular parts, then the
singular field is dualized to the worldsheet of the string and the regular field is dualized to a 2-form~\cite{Davis:1988rw, Lee:1993ty, Akhmedov:1995mw, Diamantini:1996vf, Chatterjee:2006iq, Franz:2006gb}. Let us carry out this procedure in presence of fermions which couple to the gauge field. 

We start from the action~\footnote{While it is more economical to use the form notation, the calculations in this section and the next are more transparent in the index notation.}
\begin{equation}\label{Higgs.action}
 S = \int d^4 x \left(- \frac14 F_{\mu\nu}F^{\mu\nu} +
|D_\mu\Phi|^2 
+ V(|\Phi|^2)
+ \bar{\psi}(i\slashed{\partial} -e\slashed{A} - m)\psi
\right)\,,
\end{equation}
where $\Phi$ is a complex scalar with electric charge $q$\, 
and $\psi$ is a fermion with charge $e$\,.
We have not assumed any relation between the charge $e$ of the fermion and the
charge $q$ of the scalar. The potential $V(|\Phi|^2)$ has a degenerate minimum at $\Phi^*\Phi = v^2$ for
some non-vanishing $v^2$\,, but the exact form of $V$ is not important for our calculations. 
Vortex strings or magnetic flux tubes form in this model when the global U(1) symmetry is broken in the vacuum 
and the phase of $\Phi$\,, which lives on the vacuum manifold  $\Phi^*\Phi = v^2$\,, becomes multivalued. 
The position of the ANO string is the locus of the zeroes of $\Phi$.

The corresponding partition function 
\begin{equation}
Z = \int \D A_\mu\D\Phi\D\Phi^*\D\psi\D\bar{\psi}\; \exp iS\,
\label{Higgs.Z}
\end{equation}
can be rewritten in presence of these vortex string configurations by using
a polar decomposition of the complex scalar in field space as
$\Phi = \dfrac 1{\sqrt 2}\rho\exp(i\theta)\,.$ 
Then in the presence of
flux tubes, we can write $\theta = \theta^r + \theta^s\,,$ where $\theta^s$ corresponds to a
given magnetic flux tube, and $\theta^r$ describes single valued
fluctuations around this configuration. For a string with winding number $n\,,$ 
$\theta^s$ changes by $2\pi n$ for going around the string once, while $\rho$ vanishes along
the core of the string. The ANO string world sheet $\Sigma$ is the collection of singular points~\footnote{The singular points of $\theta$ are at the zeroes of $\rho$~\cite{Lee:1993ty}.} of $\theta\,$, 
\begin{equation}\label{string-ws}
\epsilon^{\mu\nu\lambda\rho}\partial_\lambda\partial_\rho\theta = \Sigma^{\mu\nu}(x) =
2\pi n\int d^2\sigma\, \epsilon^{ab}\frac{\partial X^\mu}{\partial \sigma^a} 
\frac{\partial X^\nu}{\partial \sigma^b} \delta^4({x} - {X}(\sigma))\,,
\end{equation}
where we have included
the vorticity quantum $2\pi$ and the winding number $n$ in the definition of the world sheet. 
The string carries quantized magnetic flux,
\begin{equation}
\oint_\Gamma A_\mu dx^\mu = \frac{2n\pi}{q}\,.
\label{flux.quantum}
\end{equation}

In the polar decomposition, the action takes the form 
\begin{align}\label{polar.action}
S = \int d^4 x &\left(- \frac14 F_{\mu\nu}F^{\mu\nu} +
\frac{1}{2}\partial_{\mu}\rho \partial^\mu\rho  
+ \frac{1}{2} \rho^2 (\partial_\mu\theta^s + \partial_\mu\theta^r + q A_\mu) (\partial^\mu\theta^s + \partial^\mu\theta^r + q A^\mu)\right. \notag\\
&\qquad \left. 
+ V(\rho^2)
+ \bar{\psi}(i\slashed{\partial} -e\slashed{A} - m)\psi
\right)\,.
\end{align}
We first linearize the term $ \frac{1}{2} \rho^2 (\partial_\mu\theta + q A_\mu) (\partial^\mu\theta + q A^\mu)$ by introducing 
an auxiliary field $C_\mu$ into the partition function through the Gaussian integral 
\begin{equation}
N \int \D C_\mu\exp{\left(- \frac{i}{2}\int d^4x\, \left[{C_\mu}+ \rho(\partial_\mu \theta + qA_\mu)\right]^2\right)} = 1\,.
\end{equation}
Here $N$ is a normalization factor independent of all fields -- we will suppress $N$ and other such normalization factors for all functional integrals. We can now write the partition function as
\begin{align}
\mathcal{Z} = \int &\D A_\mu \D \rho \D \theta^s \D \theta^r \D \bar{\psi} \D \psi \D C_\mu \notag \\
 \qquad   \exp&\left(i\int d^4x \left( -\frac{1}{4}F_{\mu\nu}F^{\mu\nu} + \frac{1}{2} \partial_\mu\rho \partial^\mu\rho -\frac{1}{2}{C_\mu C^\mu}- C^\mu \rho \partial_\mu \theta^r -  C^\mu \rho (\partial_\mu \theta^s +q A_\mu)\right.\right. \notag \\
&\left. \qquad 
+ V(\rho^2) + \bar{\psi}(i\gamma^\mu \partial_\mu - m)\psi - eA_\mu \bar{\psi}\gamma^\mu\psi \bigg)\right)\,.
\end{align}
Integration over $\theta^r$ produces a $\delta(\partial_\mu(\rho C^\mu))$\,, leading to the dualization  
\begin{equation}\label{defB}
	\rho C^\mu = \frac{1}{2}\epsilon^{\mu \nu \lambda \sigma} \partial_\nu B_{\lambda\sigma}\,.
\end{equation}
Then the condition $\partial_\mu(\rho C^\mu) = 0 $ is automatically satisfied.

Putting this expression for $C^\mu$ and noting that $\varepsilon ^{\mu\nu\rho\lambda}\varepsilon _{\mu\nu_1\rho_1\lambda_1} = - \delta^{\nu\;\rho\;\lambda}_{[\nu_1\rho_1\lambda_1]}$ we get the partition function as
\begin{align} \label{Z.first}
\mathcal{Z} =& \int \D A_\mu \D \rho \D x^\mu \D \bar{\psi} \D \psi\D B_{\mu\nu} \notag\\
&     \exp\bigg(i\int d^4x \bigg( -\frac{1}{4}F_{\mu\nu}F^{\mu\nu} + \frac{1}{2} \partial_\mu \rho \partial^\mu \rho 
+\frac{1}{12\rho^2}H^{\mu\nu\lambda} H_{\mu\nu\lambda} 
- \frac{1}{2} B^{\mu\nu}\Sigma_{\mu\nu}  \notag \\
&\qquad \qquad - \frac{1}{6} q\varepsilon ^{\mu\nu\rho\lambda} A_\mu H_{\nu\rho\lambda} 
+ V(\rho^2) + \bar{\psi}(i\gamma^\mu \partial_\mu - m)\psi - eA_\mu \bar{\psi}\gamma^\mu\psi \bigg)\bigg)\,.
\end{align}
Here we have written $H_{\mu\nu\lambda} = \partial_{\mu}B_{\nu\lambda}+ \partial_{\nu} B_{\lambda\mu} + \partial_{\lambda} B_{\mu\nu}$ for the field strength of the 2-form gauge field $B$ and also replaced the integration over $\theta^s$ by an integration over the spacetime coordinates $x^\mu(\sigma, \tau)$ of the worldsheet of the ANO string. The Jacobian for this change of variables produces the action for the dynamics of the string~\cite{Akhmedov:1995mw, Polchinski:1991ax, Quevedo:1996uu, Baker:1999xn}. 

In the absence of strings we could eliminate $B_{\mu\nu}$ in favor of $A_\mu$ by summing over a perturbation series or equivalently by using the equations of motion, leaving only a massive gauge field~\cite{Cremmer:1973mg, Allen:1990gb}. We cannot do that here because of the $B_{\mu\nu}\Sigma^{\mu\nu}$ interaction. So in order to diagonalize the system and eliminate the mixed $A-B$ term, we first formally integrate over $A_\mu$\,. The relevant part of the partition function is 
\begin{align}\label{Z.A}
\int \D A_\mu  \exp i\int d^4x \bigg( -\frac{1}{4}F_{\mu\nu}F^{\mu\nu} - \frac{1}{6} q\varepsilon ^{\mu\nu\rho\lambda} A_\mu H_{\nu\rho\lambda}
- A_\mu \bar{\psi}\gamma^\mu\psi  \bigg) \,.
\end{align}
Into this we insert the following Gaussian integral 
\begin{equation}
N \int \D\chi_{\mu\nu} \exp \left(-\frac{i}{4}\int d^4 x \left\{\chi_{\mu\nu}\chi^{\mu\nu} - \varepsilon^{\mu\nu\rho\lambda} \chi_{\mu\nu} F_{\rho\lambda} + \frac{1}{4}\left(\varepsilon^{\mu\nu\rho\lambda}F_{\rho\lambda}\right)^2\right\} \right) = 1\,,
\end{equation} 
where $N$ is a constant (field-independent) normalization factor which is to lumped with other similar factors coming from other integrations.
We now integrate over $A_\mu$ to be left with 
$\delta \left(\frac{1}{2}\varepsilon^{\mu\nu\rho\lambda} \partial_\nu \left(\chi_{\rho \lambda}  
-q B_{\rho\lambda}\right)  - e \bar{\psi}\gamma^\mu\psi\right)$\,,
which can be resolved by setting 
\begin{equation}\label{chi-ansatz}
\chi_{\mu\nu} = \partial_\mu A^m_\nu - \partial_\nu A^m_\mu + q B_{\mu\nu} 
+ e \varepsilon_{\mu\nu\rho\lambda} \partial^{\rho}\frac{1}{\Box}\bar{\psi}\gamma^{\lambda}\psi\,,
\end{equation}
where we have introduced another vector field $A_\mu^m$\,, known as the magnetic photon. 
The partition function now has the form
\begin{align}\label{Z.Am}
\mathcal{Z} =& \int \D A^m_\mu\D \rho \D x^\mu \D {\psi} \D \bar\psi \D B_{\mu\nu} \notag\\
&  \exp i\int d^4x \bigg[ \frac{1}{2} \partial_\mu \rho \partial^\mu \rho +\frac{1}{12\rho^2}H^{\mu\nu\lambda} H_{\mu\nu\lambda} -\frac{1}{2} B_{\rho\lambda}\Sigma^{\rho\lambda} +
V(\rho^2) + \bar{\psi}(i\gamma^\mu \partial_\mu - m)\psi \notag\\
& -\frac{1}{4}\bigg(\partial_\mu A^m_\nu - \partial_\nu A^m_\mu + q B_{\mu\nu}\bigg)^2 -\frac{1}{2}e q B^{\mu\nu}\varepsilon_{\mu\nu\rho\lambda} \partial^{\rho}\frac{1}{\square}\bar{\psi}\gamma^{\lambda}\psi -\frac{1}{4} e^2 \bigg(\varepsilon_{\mu\nu\rho\lambda} \partial^{\rho}\frac{1}{\square}\bar{\psi}\gamma^{\lambda}\psi\bigg)^2 \bigg]\,.
\end{align}
If we now do a field redefinition $B_{\mu\nu} \to B_{\mu\nu} + \dfrac{1}{q} \left(\partial_\mu A^m_\nu - \partial_\nu A^m_\mu \right)$\,, we can integrate over $A^m_\mu$ and find that $\partial_\nu \Sigma^{\mu\nu} = 0\,,$ which shows that vortices form closed loops in this model as the world sheet current is conserved by itself~\cite{Chatterjee:2006iq}. Furthermore, we notice that the last term in the integrand above can also come from integrating over the gauge field $A_\mu$ in the partition function of ordinary quantum electrodynamics with no additional field,
\begin{equation}\label{Z.qed}
\int \D A_\mu \D {\psi} \D \bar\psi \exp i\int d^4 x \bigg( -\frac{1}{4}F_{\mu\nu}F^{\mu\nu} -\frac{1}{2\xi}(\partial_\mu A^\mu)^2 -eA_\mu\bar{\psi}\gamma^{\mu}\psi \bigg)\,.
\end{equation}
Thus we can reinstate the QED part of the action to write the Lagrangian as
\begin{equation}\label{lag}
\boxed{
\begin{aligned}
\mathscr{L} & =  -\frac{1}{4} M^2 B_{\mu\nu} B^{\mu\nu}  -\frac{1}{4}F_{\mu\nu}F^{\mu\nu} -eA_\mu\bar{\psi}\gamma^{\mu}\psi -\frac{1}{2}e M B^{\mu\nu}\varepsilon_{\mu\nu\rho\lambda} \partial^{\rho}\frac{1}{\square}\bar{\psi}\gamma^{\lambda}\psi \\
&\qquad  +\frac{1}{2} \partial_\mu \rho \partial^\mu \rho +\frac{v^2}{12\rho^2}H^{\nu\rho\lambda} H_{\nu\rho\lambda} -\frac{v}{2} B_{\rho\lambda}\Sigma^{\rho\lambda} + V(\rho^2) + \bar{\psi}(i\gamma^\mu \partial_\mu - m)\psi\,,
\end{aligned}
}
\end{equation}
where we have rescaled $B_{\mu\nu} \to v B_{\mu\nu}$\,, written $M=qv\,,$ and suppressed the gauge-fixing term. The nonlocal interaction between $B_{\mu\nu}$ and charged fermions cannot be removed as long as ANO strings are present in the system.

The first term is a mass term for the $B_{\mu\nu}$ field -- in an alternative derivation of the dual action it appears not in this form but as the equivalent nonlocal Meissner term~\cite{Mukherjee:2019vmi, Beekman2011_FOP} $\displaystyle{\frac{q^2 v^2}{12}H^{\nu\rho\lambda} \frac{1}{\square} H_{\nu\rho\lambda}}\,.$ There is an advantage of writing the mass term in the nonlocal form. The action of Eq.~(\ref{Z.first}) which was found by dualization was invariant under the vector gauge transformation $B_{\mu\nu} \to B_{\mu\nu} + \partial_{[\mu}\Lambda_{\nu]}\,.$ Written in the nonlocal form, the mass term and thus the action based on the Lagrangian of Eq.~(\ref{lag}) remains invariant under the same transformation. In this paper we will work with the mass term in the local form as in Eq.~(\ref{lag}), but we could have worked as well with the nonlocal mass term.
Another way of handling the nonlocal mass term is by introducing an additional vector field~\cite{Choudhury:2015rua}, but we will not take that route in this paper.

\section{Effective fermion interaction}\label{effpot}
The nonlocal ``spin-gauge'' coupling between the fermions and the $B$-field gives rise to a linear attractive potential between fermions, irrespective of whether they were positively or negatively charged~\cite{Chatterjee:2016liu}.
However, as compared to the action considered there, several additional terms have appeared in the derivation from the boson-fermion model. 
The effective static interaction potential between nonrelativistic fermions in this system was calculated recently~\cite{Mukherjee:2019vmi} by integrating out the gauge fields $A_\mu$ and $B_{\mu\nu}$\,. It is worthwhile to revisit this calculation, as we will be concerned in this paper with a particular modification of the result found there. 

To find the effective static potential between non-relativistic electrons, we first integrate over the gauge fields $A_\mu$ and $B_{\mu\nu}$\, using the Lagrangian of Eq.~(\ref{lag}) with $\rho=v$\,. Introducing a gauge-fixing term we can integrate over $A_\mu$\,,
\begin{align}\label{A.integral}
\int \D A_\mu \exp\, & i\int d^4x \bigg[-\frac{1}{4} F_{\mu\nu} F^{\mu\nu} -\frac{1}{2\xi} \left(\partial_\mu A^\mu\right)^2 
- eA_\mu \bar{\psi}\gamma^\mu\psi \bigg] 
\sim \exp \frac{i}{2}\int d^4 k\; J^\mu(-k) \;\frac{1}{k^2} \;J_{\mu}(k)\,,
\end{align}
where $J^\mu$ is the fermion current and we have used the fact that it is conserved, $\partial_\mu J^\mu = 0\,.$  
For the $B$ integration, we get
\begin{align}\label{B.integral}
\int \D B_{\mu\nu} \exp\, i\int d^4x  \left(-\frac{1}{4}B_{\mu\nu} K^{\mu\nu\rho\lambda}B_{\rho\lambda}
- \frac{1}{2} B_{\mu\nu}J^{\mu\nu}\right)\,,
\end{align}
where we have written  
\begin{equation}\label{B.kinetic}
K^{\mu\nu\rho\lambda} = \frac{1}{2}\left(\Box  + M^2 \right) g^{\mu[\rho}g^{\lambda]\nu}
+ \frac{1}{2}\left(g^{\nu[\rho}g^{\lambda]\sigma }\partial_\sigma \partial^\mu - g^{\mu[\rho}g^{\lambda]\sigma }\partial_\sigma \partial^\nu\right)\,,
\end{equation} 
and
\begin{equation}\label{current}
J^{\mu\nu} =v \Sigma_{\mu\nu}-e M \varepsilon_{\mu\nu\rho\lambda} \partial^{\rho}\frac{1}{\Box}\bar{\psi}\gamma^{\lambda}\psi \,.
\end{equation}
The inverse of $K^{\mu\nu\rho\lambda}$ is the propagator for $B_{\mu\nu}$ and is given in momentum space as
\begin{equation}\label{B-propagator.m}
G_{\mu\nu\rho\lambda}(k)= -\frac{1}{(k^2-M^2)}\left(g_{\mu[\rho}g_{\lambda]\nu} -\frac{1}{M^2}\left(g_{\mu[\rho}k_{\lambda]}k_\nu - g_{\nu[\rho}k_{\lambda]}k_\mu\right)\right)\,,
\end{equation}
where
\begin{equation}
K^{\mu\nu\rho\lambda} G_{\mu\nu\rho'\lambda'}= \left(\delta^\rho_{\rho'}\delta^\lambda_{\lambda'} - \delta^\rho_{\lambda'}\delta^\lambda_{\rho'}\right)\,.
\end{equation}
Thus the integration over $B_{\mu\nu}$ will result in
\begin{align}\label{B-integration.m}
& \int {\D}B_{\mu\nu}\exp\bigg( i\int d^4x \bigg(-\frac{1}{4}B_{\mu\nu}K^{\mu\nu\rho\lambda}B_{\rho\lambda}
- \frac{1}{2}B_{\mu\nu}J^{\mu\nu}\bigg)\bigg) \notag \\
&\qquad \sim \exp\bigg(\frac{i}{2}\int d^4k\, e^2\,  J^{\mu}(-k)\Bigg[\frac{1}{(k^2-M^2)} -\frac{1}{k^2}\Bigg]J^{\mu}(k)+\cdots\bigg) 
\end{align}
where we have used the expression for $J^{\mu\nu}$ in terms of the fermion current $J^\mu$. The dots stand for terms involving the string worldsheet $\Sigma$~\cite{Mukherjee:2019vmi}.

Thus we see that the $\dfrac{1}{k^2}$ appearing due to the Coulomb potential cancels with a negative term appearing after the integration of the $B$ field so that the propagator corresponds to the Yukawa potential $\dfrac{\exp(-Mr)}{r}$\,. The linear attractive potential which was found as a consequence of the nonlocal ``spin-gauge'' interaction term has disappeared.
It is because of the mass term of the $B$-field that the effective static potential between nonrelativistic fermions calculated from this action is  Yukawa. 
In a phase where the mass term is not present, the effective potential is a combination of Coulomb and linear potentials.
Let us then consider the question whether there can be such a phase.
%

\section{String Higgs mechanism and the mass of $B$}{\label{condensate}}
The photon becomes massive in the Anderson-Higgs mechanism which provides the phase transition from a massless gauge field to a massive one -- the phase transition in the opposite direction is accompanied by symmetry restoration as the system is raised above the transition temperature. 
The 2-form $B_{\mu\nu}$ is a higher analogue of the usual 1-form gauge field $A_\mu$, so what we have in mind is ``a higher analogue'' of spontaneous symmetry breaking. 
Since $B_{\mu\nu}$ couples to worldsheets of vortex strings, it is the condensation of these strings which should produce the Higgs mechanism for the $B$-field. The idea of a phase transition due to vortex condensation at finite temperature is not a new one, more generally examples are known in quantum field theory and condensed matter physics of phase transitions driven by condensation of topological defects~\cite{PhysRevLett.100.156804,2019NatCo..10.2658L,2014NatPh..10..970L, Quevedo:1996uu}.

\subsection{Higgs mechanism for $B_{\mu\nu}$}
Let us summarize the idea of a Higgs mechanism for $B_{\mu\nu}$ following~\cite{Kawai:1980qq, Seo:1981tm, Rey:1989ti, Ramos:2005yy, Forster:1978qi}. For quantized strings we can define a wavefunctional $\Psi[\Omega]$ on the space of parametrized loops $\left\{\Omega = \{X^\mu(\sigma, \tau) = X^\mu(\sigma+2\pi, \tau)\} \right\}$ which remains invariant under a reparametrization $\Omega(\sigma)\to \tilde{\Omega}(\tilde\sigma) $\,,
\begin{equation}\label{Psi-inv}
 \Psi[\tilde{\Omega}(\tilde\sigma)] = \Psi[\Omega(\sigma)]\,.
\end{equation}
Then the quantum mechanical wave equation for $\Psi$ can be written as
\begin{equation}\label{Psi-Eq}
\left[-i\frac{\delta}{\delta\Sigma^{\mu\nu}} + g B_{\mu\nu}\right]^2\Psi[\Omega] = \tau_s^2 \Psi[\Omega]\,,
\end{equation}
where $\tau_s$ is the string tension and we have written $g$ for the coupling constant between $B_{\mu\nu}$ and the worldsheet. 	
The functional derivative with respect to $\Sigma_{\mu\nu}$ is taken by making an infinitesimal keyboard deformation~\cite{Nambu:1978bd} normal to the loop at $X^\mu(\sigma, \tau)$\,. 

For the second quantized string we consider a complex scalar functional $\Psi[\Omega]$ on the loop space, invariant under reparametrizations as before. Taking a clue from the quantum mechanics of the string, we write an action of this string field interacting with the $B$-field,
\begin{equation}\label{string-Higgs-model}
S = \int d^4 x \frac{1}{12} H_{\mu\nu\lambda}^2 - \int [d x(\cdot)]\oint \sqrt{h} d\sigma \left[\left|\frac{\delta\Psi[\Omega]}{\delta\Sigma^{\mu\nu}} + i g B_{\mu\nu}\Psi[\Omega]\right|^2 + \mu^2|\Psi[\Omega]|^2\right] + S_{\rm int}\,.
\end{equation}
The string interaction term $S_{\rm int}$ represents splitting and joining of strings through cubic, quartic and similar terms. There should be interaction terms between the string and the Higgs field as well. This action is invariant under a global U(1) transformation $\Psi[\Omega] \to e^{i\omega}\Psi[\Omega]$ and also under string reparametrizations $\Psi[\Omega(\sigma)] \to \tilde{\Psi}[\tilde{\Omega}(\tilde{\sigma})]\,.$  Following~\cite{Marshall:1974wf}, we gauge the U(1) transformation by making it local on loop space,
\begin{equation}\label{string-gauge-transf}
\Psi[\Omega] \to \Psi'[\Omega] = e^{ig\omega[\Omega]}\Psi[\Omega]\,, \qquad \omega[\Omega(\cdot)] = \oint_\Omega dx^\mu \Lambda_\mu\,.
\end{equation}
Then the gauge covariant area derivative becomes
\begin{equation}\label{B-connection}
\frac{\delta\Psi'}{\delta\Sigma^{\mu\nu}} + i g B'_{\mu\nu}\Psi' = e^{i\omega[\Omega]}\left[\frac{\delta}{\delta\Sigma^{\mu\nu}} + i g \left(B'_{\mu\nu} - \frac{1}{g}\partial_{[\mu}\Lambda_{\nu]}\right)\right]\Psi\,.
\end{equation}
Thus the gauge covariant derivative transforms homogeneously and the action remains invariant provided $B_{\mu\nu}$ undergoes a vector gauge transformation as $B_{\mu\nu}\to B'_{\mu\nu} = B_{\mu\nu} + \partial_{[\mu}\Lambda_{\nu]}\,.$ 

The idea of a Higgs mechanism corresponding to this symmetry is a generalization of the usual one for Abelian gauge fields. The vortex loops condense into the vacuum for $\mu^2 < 0$ and the functional field gets a nonvanishing vacuum expectation value (vev). The simplest case is when this vev is a constant in spacetime, i.e. $\langle 0 | \Psi[\Omega] | 0\rangle = \frac{1}{2g} M_B\,,$ in which case the Lagrangian for the ``free'' $B$-field becomes
\begin{equation}\label{B-lag-vev}
{\mathscr L} = \frac{1}{12} H^2_{\mu\nu\lambda} - \frac{1}{4} M_B^2 B^2_{\mu\nu}\,.
\end{equation}
This is thus exactly like the Higgs mechanism, but the gauge field which becomes massive is $B_{\mu\nu}$\,. We will call this the Higgs mechanism for strings, and distinguish it from the ``usual Higgs mechanism'', by which we will mean the Higgs field $\Phi$ getting a nonvanishing vev and the photon $A_\mu$ becoming massive.  

In the Higgs mechanism for strings, the mass of the $B_{\mu\nu}$ field is generated by the condensate of vortex loops. In analogy with the usual Higgs mechanism, we think of the state where loops have condensed as the ``true vacuum''. This state should be the global minimum of the energy in which both the Higgs field $\Phi$ and the string condensate field $\Psi[\Omega]$ are frozen at their respective nonvanishing vevs. If there is a Higgs mechanism for strings at work, there should be another state, a ``false vacuum''  in which $\langle\Psi[\Omega]\rangle = 0$\, and the $B_{\mu\nu}$ field is massless. This false vacuum state is clearly not a local minimum of the energy, so (semi)classical strings are unstable. In particular, the dualization procedure of Sec.~\ref{dual-string} can be done only in the true vacuum state where the strings are frozen, i.e., fluctuations due to creation and annihilation of strings can be ignored. In general, the couplings and masses differ from their values at the true vacuum by quantum corrections\footnote{For example, we cannot write $M = qv\,.$}. The question now is whether a false vacuum state exists for $\Psi$\,, to which the answer is known, at least for the simpler theory with no fermions.
%

\subsection{False vacuum for $\Psi[\Omega]$}
In order to understand the phases of the system of vortex strings, in particular whether it may have a false vacuum, we calculate the Euclidean partition function~\cite{Forster:1978qi, Seo:1981tm, Kawai:1980qq, Rey:1989ti, Ramos:2005yy}. Let us consider a hypercubic lattice in four Euclidean dimensions, with lattice spacing $a$\,. The $B_{\mu\nu}$ field couples to area elements, so it is defined in terms of the plaquette operator on a plaquette $p$ as 
\begin{equation}\label{plaquette}
U_p(B_{\mu\nu}) = \exp [-i g a^2 B_{\mu\nu}(p)]\,.
\end{equation}
The vector gauge transformation $B_{\mu\nu} \to B_{\mu\nu} + \partial_{[\mu}\Lambda_{\nu]}\,$ acts along the links on the boundary of a given plaquette $p$\,. Writing $\Lambda_l = \exp[-iga\Lambda_\mu]$ we find that the plaquette operator transforms under the gauge transformation as 
\begin{equation}\label{plaquette-transf}
U_p(B_{\mu\nu}) \to \left[\prod_{l\in \partial p} \Lambda_l \right] U_p(B_{\mu\nu})\,.
\end{equation}
Then the gauge invariant kinetic term in Eq.~(\ref{string-Higgs-model}) is the sum over all lattice cubes of the product of the plaquette operators residing on the boundary of each cube,
\begin{equation}\label{lattice-kinetic}
\int d^4 x \frac{1}{12} H^2 = \beta \sum_{\rm cube} {\rm Re} \left[ \prod_{p\in \partial({\rm cube})} U_p\right]\,,
\end{equation}
where the lattice coupling constant of the $B_{\mu\nu}$ field has been denoted by $\beta$\,. 

For the ``kinetic term'' of the functional field $\Psi[\Omega]$ the effective Lagrangian is calculated from a sum over configurations
\begin{equation}\label{config-sum}
K(C_1, C_2, A) = e^{-\tau_s a^2 A}\sum_S \left[\prod_{p\in S} U_p(B_{\mu\nu})\right]\,,
\end{equation}
where the sum is over all Euclidean world sheets $S$ of area $A$ connecting the closed curves $C_1$ and $C_2$\,, the bare string tension is $\tau_s$\,, and the surface $S$ is taken to be orientable and without holes. 

Modifying $C_2$ by a keyboard-like plaquette variation at a link produces a recursion relation
\begin{equation}\label{recursion}
K(C_1, C_2, A) = \sum_p\left[\bar{U}_p K(C_1, C_2 + p, A - a^2) + U_p K(C_1, C_2 - p, A - a^2)\right]\,,
\end{equation}
where this sum is over all plaquettes which can be added at the given link. Since there are $2(d-1)$ such plaquettes in $d$ dimensions, it follows that $K(C_1, C_2, A)$ satisfies the diffusion equation
\begin{equation}\label{diffusion}
\frac{\partial}{\partial  \bar{A}} K(C_1, C_2, \bar{A}) = \left[\sum_p \left(\frac{\delta}{\delta\Sigma^{\mu\nu}} + i g B_{\mu\nu}\right)^2 - M_0^2\right]  
K(C_1, C_2, \bar{A})\,,
\end{equation}
where we have written $\bar{A} = a^2 A e^{-\tau_s a^2}$ and the dynamical string tension $M_0$ is related to the bare string tension $\tau_s$ as
\begin{equation}\label{string-tension}
 M_0^2 = \frac{1}{a^4}(e^{\tau_s a^2} - 6)\,
\end{equation} 
in four spacetime dimensions. The propagator $G(C_1, C_2)$ of a closed string can be written as 
\begin{equation}\label{string-propagator}
G(C_1, C_2) = \int\limits_0^\infty d\bar{A}\, K(C_1, C_2, \bar{A})\,,
\end{equation}
which can be obtained from the action
\begin{equation}\label{string-field-action}
S(\Psi[\Omega], B_{\mu\nu}) =  \int [d x(\cdot)]\oint \sqrt{h} d\sigma \left[\left|\frac{\delta\Psi[\Omega]}{\delta\Sigma^{\mu\nu}} + i g B_{\mu\nu}\Psi[\Omega]\right|^2 - M_0^2|\Psi[\Omega]|^2\right]\,.
\end{equation}
This is the same as the second term of Eq.~(\ref{string-Higgs-model}) written in Euclidean space and with $\mu^2 = M_0^2\,.$ Thus we see that it is possible that in the action Eq.~(\ref{string-Higgs-model}) of the vortices, we may have $\mu^2 <0$ for some system of the kind we are considering. The effective potential for $\Psi$ will in general have contributions also from its interaction with $\Phi^\dagger\Phi$ and with the fermions, which will modify the expression for $M_0^2$\,. The first kind of interaction appears to involve  $(\Phi^\dagger\Phi)^{-1}$\, when the strings are close to being stable, while the fermion interaction is nonlocal in the leading order. 
%

\subsection{Linear potential}
A negative $\mu^2$ for  $\Psi[\Omega]$ is not an unambiguous indication of a phase transition for vortex strings in every system where they form, as quantum corrections can change the sign of $\mu^2$\,. However we can expect that in some systems there will be a false vacuum for strings and thus a second phase transition, for some critical values of the coupling constants and the temperature. In such cases, when $\Psi[\Omega]$ is in the false vacuum with a vanishing vev, the ``free'' $B$ field is massless, i.e. in the absence of interactions its dynamics should be described by Eq.~(\ref{B-lag-vev}) with $M_B=0\,$. Then after interactions are included, the dynamics should be described by Eq.~(\ref{lag}), but without the mass term $M^2 B_{\mu\nu}B^{\mu\nu}$\,. Furthermore, while quantum corrections will change the coefficients of all the terms, there is no reason why equal coefficients at the true vacuum should remain equal after quantum corrections, except when the relations among such corrections are fixed by symmetry. Therefore we can write the effective Lagrangian as 
\begin{equation}\label{lag-final}
	\boxed{
		\begin{aligned}
			\mathscr{L} & =   -\frac{1}{4}F_{\mu\nu}F^{\mu\nu} 	+\frac{1}{12}H^{\nu\rho\lambda} H_{\nu\rho\lambda} +  \bar{\psi}(i \slashed\partial - m)\psi  \\
			&\qquad  
		 -e\bar{\psi}\slashed{A}\psi -\frac{1}{2} \tilde{g} B^{\mu\nu}\varepsilon_{\mu\nu\rho\lambda} \partial^{\rho}\frac{1}{\square}\bar{\psi}\gamma^{\lambda}\psi -\frac{1}{2}g_s B_{\rho\lambda}\Sigma^{\rho\lambda}\,,
		\end{aligned}
	}
\end{equation}
where the fields and their couplings now include quantum corrections. 
We have assumed that the mass of $\rho$ remains much bigger than the energies of thermal fluctuations, so in particular $\rho$ is frozen. This is necessary for the system to remain in the state where strings can form.

We calculate the effective static potential between a pair of charged fermions by integrating out the gauge fields from the action as before, but now we will need a gauge-fixing term $\frac{1}{2\eta}(\partial_\nu B^{\mu\nu})^2$ because the $B$-field is massless. Then the $B$-propagator is
\begin{equation}\label{B-propagator.0}
G^{\mu\nu\rho\lambda}(k)= -\frac{1}{k^2}\left(g_{\mu[\rho}g_{\lambda]\nu} -\frac{1 - \eta}{k^2}\left(g_{\mu[\rho}k_{\lambda]}k_\nu - g_{\nu[\rho}k_{\lambda]}k_\mu\right)\right)\,.
\end{equation}
Using this in the $B$-integration of Eq.~(\ref{B-integration.m}), but with the coupling constants as in Eq.~(\ref{lag-final}), we get 
\begin{align}\label{B-integration.0}
 \int {\D}B_{\mu\nu}\exp & \bigg(i\int d^4x \bigg(-\frac{1}{4}B_{\mu\nu}K^{\mu\nu\rho\lambda}_0 B_{\rho\lambda}
- \frac{1}{2}B_{\mu\nu}J^{\mu\nu}\bigg)\bigg) \notag \\
& \qquad \sim \exp\bigg( \frac{i\tilde{g}^2}{2}\int d^4k\;  J^{\mu}(-k)\frac{1}{k^4}J^{\mu}(k)+ \cdots \bigg)\,,
\end{align}
where we have written $K^{\mu\nu\rho\lambda}_0 $ for the kinetic operator with $M=0$\,,
\begin{equation}\label{B.Kinetic0}
K^{\mu\nu\rho\lambda}_0 = \frac{1}{2} g^{\mu[\rho}g^{\lambda]\nu} \Box 
	+ \frac{1}{2}\left(1 - \frac{1}{\eta}\right)\left(g^{\nu[\rho}g^{\lambda]\sigma }\partial_\sigma \partial^\mu - g^{\mu[\rho}g^{\lambda]\sigma }\partial_\sigma \partial^\nu\right)\,.
\end{equation}
The effective static potential between electrons is then attractive and linear,
\begin{equation}\label{linear}
V(r) = \tilde{g}^2 r\,.
\end{equation}
This suggests that the electrons are joined by a string.

\section{Discussion}
It is well known that the Abelian Higgs model with flux strings has a dual description in terms of a two-form potential (sometimes called the disorder field in the context of superconducting phase transitions~\cite{Kleinert:1989kx, Kiometzis:1995eg, Kiometzis:1995oku}). We have shown that the dual field has an effective nonlocal interaction with electrons in the system. The description of flux strings in terms of the Abelian Higgs model is an effective one for real type-II superconductors. The string interactions, as well as quantities like string tension or thickness, are determined by the properties of the underlying real system. The effective theory of infinitely thin flux strings in Abelian Higgs model cannot tell us what the terms in the string potential should be. The false vacuum for the string field may even be different from the false vacuum of the scalar field. However, we can say that the mass term for the $B$ field will not be present in the action in the false vacuum.
We also see that the remaining terms which contain the $B_{\mu\nu}$ field have an enhanced symmetry -- the vector gauge symmetry $B_{\mu\nu} \to B_{\mu\nu}+ \partial_\mu \Lambda_\nu - \partial_\nu \Lambda_\mu\,.$ While quantum corrections cannot be calculated without knowledge of the exact form of the string field potential in Eq.~(\ref{string-Higgs-model}), this symmetry does not fix any relation among the parameters in the false vacuum, so those terms need not vanish.
Thus if the flux strings condense in a stringy Higgs mechanism, there is a phase in which the static potential between non-relativistic electrons has a linear component.

\begin{figure}[t]
	\begin{center}
	\begin{subfigure}{}
	\includegraphics[scale=0.3]{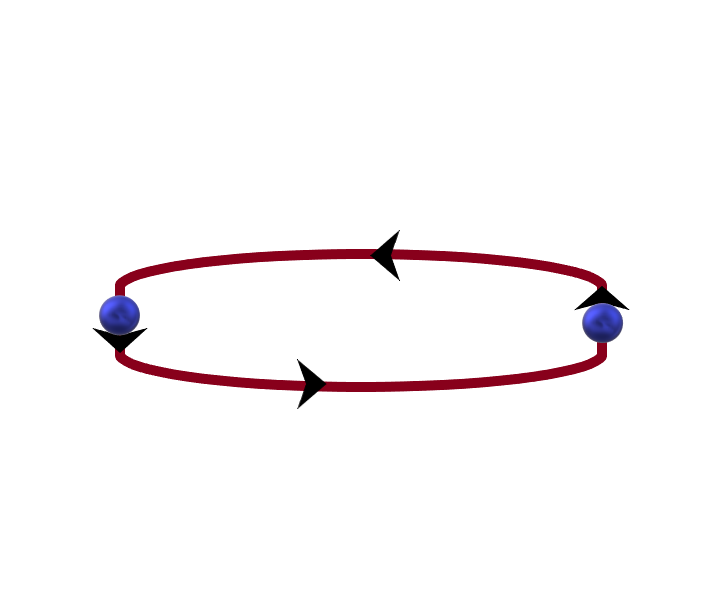}
	\end{subfigure}
\begin{subfigure}{}
	\includegraphics[scale=0.3]{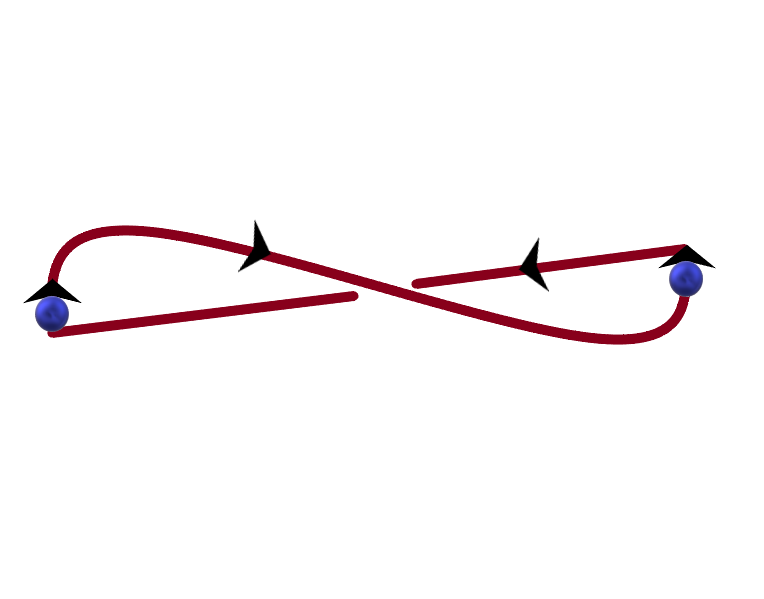}
	\end{subfigure}
	\caption{A pair of electrons connected by flux tubes. The arrows indicate the direction of magnetic flux.}	\label{string-pair}
	\end{center}
\end{figure}

\begin{figure}[t]
	\begin{center}
	\includegraphics[scale=0.3]{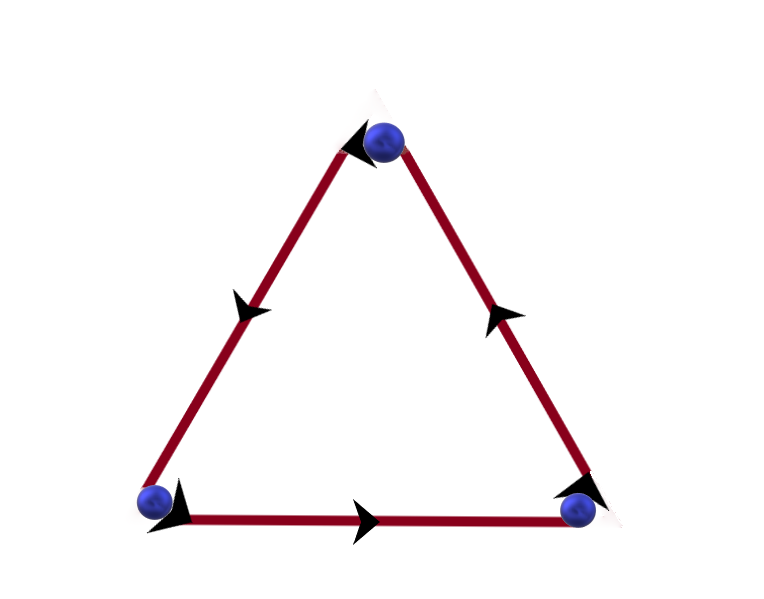}
	\caption{Three electrons connected by magnetic flux tubes.}
		\label{triangle}
	\end{center}
\end{figure}

A linear potential between pairs of particles would be interpreted as a flux tube. But there is a problem -- the flux tubes we started out with are either infinite or end on magnetic charges~\cite{Chatterjee:2006iq}, not electric charges. A possible solution to this problem comes from noting that the nonlocal interaction couples the 2-form field with the electron spin current, more precisely, the magnetic dipole moment of electron. In the background where flux lines coalesce into flux tubes, a pair of flux tubes can connect two dipoles and form a kind of Cooper pair. A cartoon representing a semiclassical view of the electron pair is shown in Fig.~\ref{string-pair}, the north pole of each dipole connected to the south pole of the other by a flux tube. In our construction the dipoles are electrons, thus point dipoles, and the flux tubes are infinitesimally thin. The dipoles can be antiparallel or parallel as in the figure, or they can be oriented in any which way with respect to each other. 
Recently a model of superinsulators was proposed using Cooper pairs bound by strings of electric flux~\cite{Diamantini:2019wbf, Diamantini:2018mjg, Diamantini:2020sej}. Our construction here is different from that one in two important aspects. One is that electric flux strings appear in dual superconductivity, which exhibits a dual Meissner effect and excludes electric fields from the bulk, constricting the flux into the strings (in analogy to QCD confinement~\cite{Polyakov:1996nc, tHooft:1999cgx, Ripka:2003vv, tHooft:2003lzk}). 
In our construction, we have the usual (type II) superconductivity arising from condensation of electric charges, which constricts the magnetic field into strings. The other difference is that in that model, electrically charged Cooper pairs are connected by the strings, the electric flux in a string ends on electric charges at the ends. In our construction, electrically charged fermions are connected by strings carrying magnetic flux, because charged fermions behave as point magnetic dipoles due to their spin. The magnetic field of a point dipole is constricted into a pair of strings in a superconductor, connecting it to another fermion as in Fig.~\ref{string-pair}. Bound states of three or more electrons are also possible, as shown in the example of Fig.~\ref{triangle}. 

We can estimate the energy of a localized pair as a function of its angular momentum, if we pretend that the electrostatic potential between two electrons joined by a magnetic flux tube remains the usual Coulomb potential. For a nonrelativistic string of length $L$ and tension $T$\,, with electrons of mass $m$ and charge $e$ at the two ends, the energy is then 
\begin{equation}\label{energy}
	E = 2m + TL + \frac{1}{2} I\omega^2 + \frac{e^2}{L}\,,
\end{equation}
where $I$ is the moment of inertia,
\begin{equation}\label{mom-inertia}
	I = \frac{1}{12} TL^3 + \frac{1}{2}mL^2\,.
\end{equation}
Writing $J=I\omega$ for the angular momentum, we find the relation between the energy $E$ and the angular momentum $J$ of the string 
\begin{equation}\label{E-J} 
	E = 2m +TL +\frac{6J^2}{L^2(TL + 6m)} + \frac{e^2}{L} = 2m + E_{b}\,,
\end{equation}
where we have defined $E_{b}$ to be the binding energy.
The same result can also be found by taking the non-relativistic limit of the relativistic string with massive end points~\cite{Sonnenschein:2018aqf}.
\begin{figure}[hbtp]
	\begin{center}
		\includegraphics[scale=0.5]{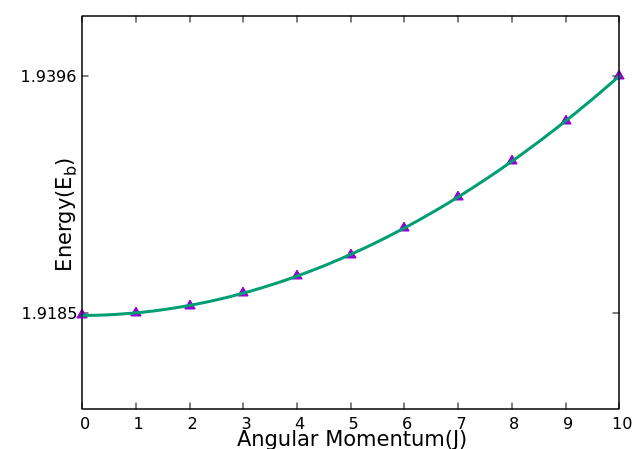}
		\caption{$E_{b}$ (in eV) vs $J$ (in $\hbar$) plot }
			\label{regge}
	\end{center}
\end{figure}
Minimizing the energy with respect to the length, one can calculate $L$ as a function of the other parameters. Putting this back into the equation for $E$\, we get a relation between the energy and the angular momentum. For type II superconductors the penetration depth is of the order $\sim$100~nm, which gives a representative string tension $T \sim 10$ eV$^2$. The corresponding plot of $E_{b}$ vs $J$ is shown in Fig.~\ref{regge}\,. The size of the pair corresponding to the $J=0$ state is $\sim$19~nm.
The energies are higher for bound states of more electrons -- for example, $E_{b} = \frac{3}{2} TL + \frac{3e^2}{L}$ for the $J=0$ state of three electrons. 

The possibility of having different numbers of fermions in bound states, with different geometries, suggests that our construction can also be useful as a toy model of quark confinement. 
In the usual string picture of quark confinement~\cite{tHooft:1999cgx, tHooft:2003lzk, Ripka:2003vv}, the vacuum is thought of as a dual (color) superconducting vacuum in which magnetic monopoles condense, dual Meissner effect takes place, and (chromo)electric flux tubes end on quarks carrying (chromo)electric charge. Here we have found another description -- that of magnetic dipoles, rather than monopoles, being connected by strings carrying magnetic flux. The electric charge of the fermion is screened by virtue of being in a superconductor.

\section*{Acknowledgements}
We acknowledge discussions with C. Chatterjee and I. Dutta Choudhury during the early stages of this work. AL acknowledges discussions with J.~K.~Bhattacharjee.

\end{document}